\documentclass{IOS-Book-Article}

\usepackage{mathptmx}
\usepackage{soul}\setuldepth{article}

\usepackage{graphicx}
\usepackage{hyperref}

\hypersetup{
    colorlinks=true,
    linkcolor=blue,
    filecolor=blue,      
    urlcolor=blue,
    citecolor=blue,
    pdfpagemode=FullScreen,
}

\usepackage{booktabs}% http://ctan.org/pkg/booktabs
\usepackage{xurl} % breaks urls

\usepackage{listings}
\usepackage{xcolor}
\usepackage{amssymb,amsmath,amsfonts}
\usepackage{censor}

\definecolor{codegreen}{rgb}{0,0.6,0}
\definecolor{codegray}{rgb}{0.5,0.5,0.5}
\definecolor{codepurple}{rgb}{0.58,0,0.82}
\definecolor{backcolour}{rgb}{0.95,0.95,0.92}
\definecolor{eclipseBlue}{RGB}{42,0.0,255}
\definecolor{eclipseGreen}{RGB}{63,127,95}
\definecolor{eclipsePurple}{RGB}{127,0,85}
\definecolor{codeyellow}{RGB}{204,122,0}

\lstdefinelanguage{SPARQL}
{
  morekeywords={
    SELECT,
    FILTER,
    WHERE,
    PREFIX,
    MINUS,
    ORDER,
    BY,
    CONTAINS,
    ASC
  },
  sensitive=false, % keywords are not case-sensitive
  morestring=[b]" % defines that strings are enclosed in double quotes
}
\lstdefinestyle{mystyle}{
	language={SPARQL},
    backgroundcolor=\color{backcolour},   
    keywordstyle=\color{codegreen},
    columns=fixed, %make all columns equal width
    numberstyle=\color{codegray},
    stringstyle=\color{codepurple},
    basicstyle=\ttfamily\footnotesize,
    breakatwhitespace=false,         
    breaklines=true,                 
    captionpos=t,                    
    keepspaces=true,                 
    numbers=left,                    
    numbersep=7pt,                  
    showspaces=false,                
    showstringspaces=false,
    showtabs=false,                  
    tabsize=2
}
\usepackage{tikz}
\usepackage{amsmath}

\def\hb{\hbox to 11.5 cm{}}

\begin{document}
%\pagenumbering{arabic}
\newcommand{\name}{ATC-Sense}

\pagestyle{plain}

\begin{frontmatter}              % The preamble begins here.

%\pretitle{Pretitle}
\title{Ontology-Based Anomaly Detection for Air Traffic Control Systems}

%\markboth{}{June 2021\hb}
%\subtitle{Subtitle}

\author[A]{\fnms{Christopher} \snm{Neal}}, %
%\thanks{Corresponding Author: christopher.neal@polymtl.ca}},
\author[A]{\fnms{Jean-Yves} \snm{De Miceli}}, 
\author[B]{\fnms{David} \snm{Barrera}}, and
\author[A]{\fnms{José} \snm{Fernandez}}

\runningauthor{Neal et al.}
\runningtitle{Ontology-Based Anomaly Detection for Air Traffic Control Systems}

%\runningauthor{\thepage}
%\runningtitle{Ontology-Based Anomaly Detection for Air Traffic Control Systems}
\address[A]{Polytechnique Montreal, Canada}
\address[B]{Carleton University, Canada}

\begin{abstract}
The Automatic Dependent Surveillance-Broadcast (ADS-B) protocol is increasingly being adopted by the aviation industry as a method for aircraft to relay their position to Air Traffic Control (ATC) monitoring systems. ADS-B provides greater precision compared to traditional radar-based technologies, however, it was designed without any encryption or authentication mechanisms and has been shown to be susceptible to spoofing attacks. A capable attacker can transmit falsified ADS-B messages with the intent of causing false information to be shown on ATC displays and threaten the safety of air traffic. Updating the ADS-B protocol will be a lengthy process, therefore, there is a need for systems to detect anomalous ADS-B communications. This paper presents \name, an ADS-B anomaly detection system based on ontologies. An ATC ontology is used to model entities in a simulated controlled airspace and is used to detect falsified ADS-B messages by verifying that the entities conform to aviation constraints related to aircraft flight tracks, radar readings, and flight reports. We evaluate the computational performance of the proposed constraints-based detection approach with several ADS-B attack scenarios in a simulated ATC environment. We demonstrate how ontologies can be used for anomaly detection in a real-time environment and call for future work to investigate ways to improve the computational performance of such an approach. 
\end{abstract}

\begin{keyword}
Ontologies\sep Anomaly Detection \sep Automatic Dependent Surveillance-Broadcast \sep Air Traffic  Control

\end{keyword}
\end{frontmatter}
%\markboth{June 2021\hb}{June 2021\hb}
%\thispagestyle{empty}
%\pagestyle{empty}

\section{Introduction}\label{sec:intro}

%1.ATC
%define ATC and ATC controller
Air Traffic Control (ATC) is a service provided to ensure the flow of air traffic in a controlled airspace occurs in a safe and efficient manner~\cite{AirlineIndustry}. An air traffic controller (or simply \textit{controller}) has the goal of ensuring separation between aircraft operating in an airspace. The controller provides instructions to pilots over radio communications to do things such as change an aircraft's speed, altitude, and direction. 

As aircraft navigate through a controlled airspace, several types of surveillance technologies are used to identify individual aircraft and generate position reports. This information is used to populate the displays used by a controller to see the position of aircraft in real-time. Traditionally, this has been done using Primary Surveillance Radar (PSR) and Secondary Surveillance Radar (SSR) which broadcast signals over a geographic area and receive a response from aircraft within their range. This approach is gradually being upgraded with Automatic Dependent Surveillance-Broadcast (ADS-B) systems which can more precisely identify aircraft and their properties. In the ADS-B paradigm, an aircraft determines its position using a Global Positioning System (GPS) and broadcasts its position, speed, altitude, and flight number using digital ADS-B communication packets to ADS-B antenna receivers. ADS-B is a central component of the Next Generation Air Transportation System in the United States~\cite{NextGenADSB} as well as other air traffic surveillance modernization projects across the globe. ADS-B is also currently being used in regions which traditionally had no radar coverage, such as the Canadian Arctic~\cite{adsbNavCanada}.

%3.
The ADS-B protocol was developed with the focus on performance and not on security. By design, there are no mechanisms for authentication or encryption. This had not been a problem until the early 2000s with the emergence of low-cost and performant tools, such as Software Defined Radios (SDRs), which can be used to broadcast falsified signals to ADS-B antennas~\cite{Strohmeier2017}. An SDR device can perform complex signal processing tasks using a general-purpose computer processor, instead of requiring specialized hardware components not historically available to the general public. The feasibility of attacks which target the ADS-B communication protocol has been shown to be theoretically attainable by both academic~\cite{Costin} and hacking communities~\cite{Schafer}.

Due to the size of the airline industry, it is challenging to coordinate the large number of stakeholders and parties to resolve the security issues inherent to the ADS-B protocol in the short term. While the industry converges on new security standards, there is a gap to be filled by external systems which can be used to detect attacks on ATC systems. This paper proposes a system, called \name, to detect falsified ADS-B messages. The developed system operates in a simulated ATC environment where realistic communications data is generated by aircraft as they navigate through a controlled airspace. Attacks are injected into the simulation in-line with the capabilities of an attacker which can exploit the unsecured ADS-B protocol. Through the use of ontologies (a method for modeling information as a knowledge graph), constraints-based detection algorithms based on semantic reasoning are used to infer the presence of malicious communications traffic. \name~is the first attempt, to our knowledge, to create an ontology-based method for detecting ADS-B attacks in a real-time simulation.

%a platform that uses ontology-based detection of ADS-B attacked in a real-time simulation of a controlled airspace.

%5.Outline
The remainder of the paper is organized as follows. Section~\ref{sec:related} provides a background to the components used in ATC, the current status of ADS-B anomaly detection, and an overview of how ontologies can be used to detect attacks. Section~\ref{sec:system_overview} provides an overview of \name~and its detection approach. A description and evaluation of the \name~detection approach is provided in Section~\ref{sec:evaluation}. We conclude with Section~\ref{sec:conclusion}.

%-- Section 2: Background and Related work
\section{Background and Related Work}\label{sec:related}
%This section provides the reader with an overview of ATC concepts and the security challenges ATC systems face. Additionally, a review of anomalous ADS-B detection approaches is provided along with an outline of ontology-based anomaly detection techniques.
This section provides an overview of core ATC concepts and security challenges, a review of anomalous ADS-B detection approaches, and an outline of ontology-based anomaly detection processes.

\subsection{ATC Components and Security Issues}

%\subsubsection{Flight Plan.}
\paragraph{Flight Plan.} Prior to a flight, a pilot must submit a flight plan to all the agencies which control an airspace wherein the aircraft will travel. A flight plan contains information such as the origin airport, destination airport, a planned route, alternate routes, aircraft information, aircraft type, etc. When an aircraft enters a controlled airspace, a flight strip is created and assigned to a controller to track the~flight. 

\paragraph{Position Reporters.} As aircraft travel through an airspace, timestamped position reports are generated by position reporter systems that identify the latitude, longitude, angle, and potentially other information about the aircraft. The list of reported positions for an individual aircraft constitute the flight's track. These are generally generated by ground-based stations with a limited coverage area. This paper focuses on PSR, SSR, and ADS-B position reporters.

%PSR
\paragraph{Primary Surveillance Radar (PSR).}PSR works by echoing signals off of physical objects and does not require any specialized equipment in an aircraft. These types of signals cannot be falsified since they work on physical properties, however, they provide limited information. PSR can identify the longitude and latitude of aircraft, but it cannot determine its altitude or uniquely identify the aircraft. %PSR is still used in ATC systems today, however, in a complementary nature to more sophisticated technologies.

%SSR
\paragraph{Secondary Surveillance Radar (SSR).}To overcome the shortcomings of PSR, SSR was developed to uniquely identify aircraft through the use of a \textit{transponder}. A Transponder is an electronic device in an aircraft that provides a coded signal in response to an interrogating device. A transponder transmits an aircraft's position, altitude, and identity. SSRs are generally mounted on top of PSR installations. SSR signals can potentially be spoofed by an attacker, however, it is extremely difficult and is not the focus of this paper. The difficulty lies in the requirement of continuously spoofing the handshake-like operation in accordance with the physics of the SSR groundstation capabilities. 
  
%SSR broadcasts are susceptible to eavesdropping, however, it is unlikely SSR messages can be readily spoofed

%This is possible because there is no authentication or encryption on SSR messages. %SSR technologies are the primary method for tracking aircraft in ATC today.

\begin{figure}[h!]
\centering
	\includegraphics[scale=0.25]{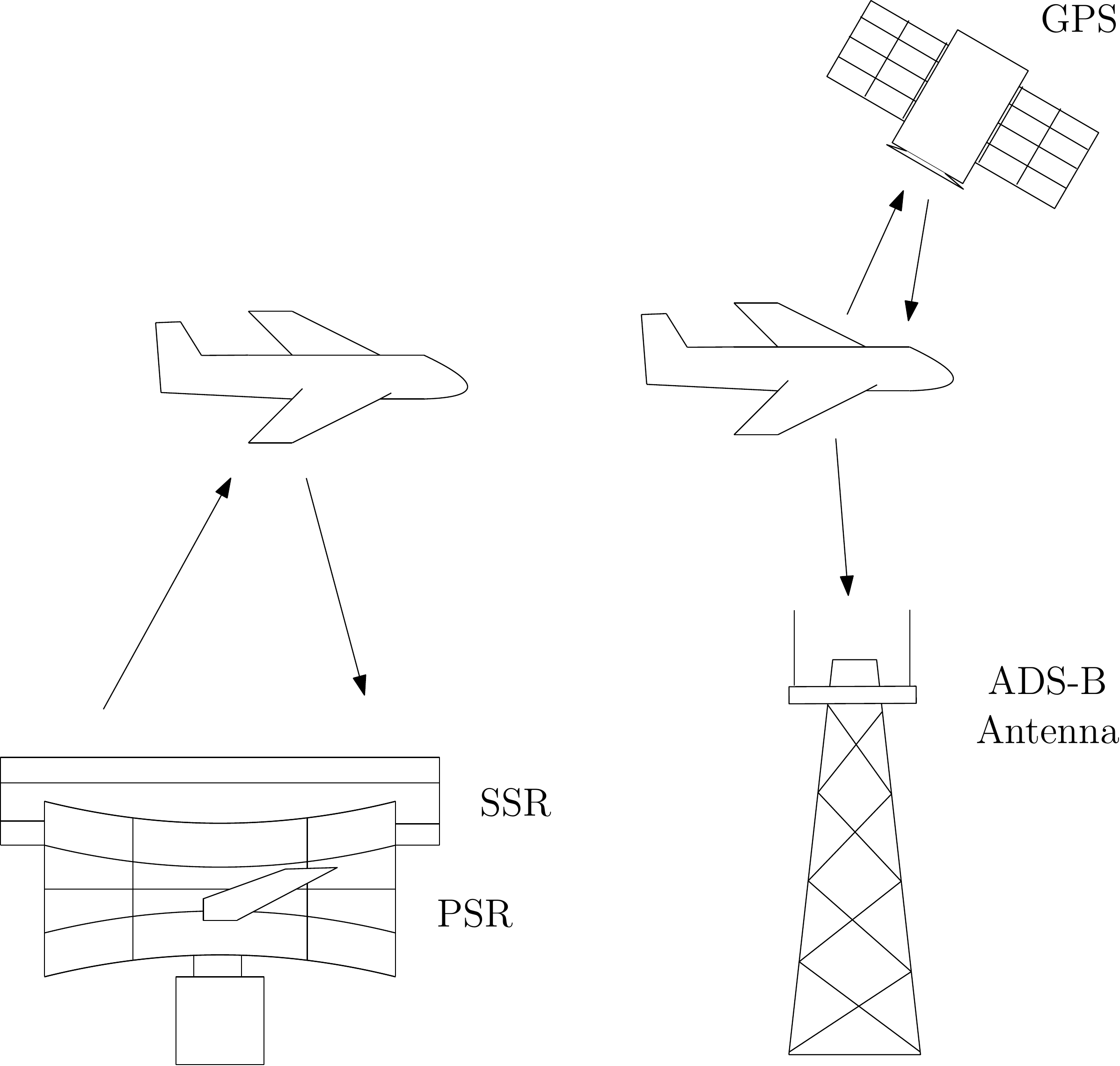}
	\caption{PSR, SSR, and ADS-B reporting principles. PSR and SSR systems send signals over a geographic area and receive a response from aircraft in the area. ADS-B antennas receive signals from aircraft that have determined their position via GPS.}
	\label{fig:report_techs}
\end{figure}

%ADS-B
\paragraph{Automatic Dependent Surveillance-Broadcast (ADS-B).} In the ADS-B paradigm, an aircraft determines its position via GPS and broadcasts this to ADS-B antennas as well as other aircraft over SSR-type communications using an aircraft's transponder. ADS-B is advantageous as it provides more precise tracking of aircraft, however, this technology was developed with a focus on performance and not security. There is the possibility for attacks which spoof ADS-B signals to ground-based antennas~\cite{Schafer,Costin}. The core concepts of PSR, SSR, and ADS-B technologies are visualized in Figure~\ref{fig:report_techs}.

%ADS-B is a more recent development and is gaining in adoption as it offers more accurate position reporting through the use of GPS. The idea is that an aircraft can determine its position via GPS and broadcast it to ground ADS-B antennas as well as other aircraft over SSR-type communications using an aircraft's transponder. ADS-B has several advantages as it provides more precise tracking of aircraft and it will eventually provide the ability for pilots to see the location of ADS-B enabled aircraft on a visual display similar to what an air traffic controller has access to. However, this technology was developed with a focus on performance and not security. There is the possibility for attacks which spoof ADS-B signals to ground-based antennas~\cite{Schafer,Costin}. The core concepts of PSR, SSR, and ADS-B technologies are visualized in Figure~\ref{fig:report_techs}

%There is the possibility for GPS-based attacks as well as the spoofing of ADS-B signals between aircraft transponders and antennas~\cite{Schafer,Costin}. 
% and summarized in Table~\ref{tab:atc_comm_techs}.

\iffalse
\begin{table*}[t]
  \centering
  \caption{Overview of PSR, SSR, and ADS-B}
\begin{tabular}{|p{1.1cm}|p{3.5cm}|p{3.5cm}|p{3.5cm}|}
  \hline
  \multicolumn{4}{|c|}{\textbf{ATC Reporting Technologies}}\\
  \hline
  \hline
	\multicolumn{1}{|c|}{\small{\textbf{Tech.}}} & \multicolumn{1}{|c|}{\small{\textbf{Reporting Principle}}} & \multicolumn{1}{|c|}{\textbf{\small{Information Captured}}} & \multicolumn{1}{|c|}{\textbf{\small{Security Concerns}}}\\
  \hline
	\multicolumn{1}{|c|}{\small{PSR}} & \small{Groundstation broadcasts electromagnetic waves and judges reflection off of physical objects} & \small{Aircraft distance, angle, and radial velocity} & \small{Cannot be falsified, susceptible to physical noise}\\
	\hline
	\multicolumn{1}{|c|}{\small{SSR}} & \small{Groundstation broadcasts radio interrogations and receives encoded radio response from aircraft transponder} & \small{Aircraft position, altitude, and identity} & \small{Susceptible to eavesdropping and falsification (higher degree of difficulty)}\\ 
	\hline
	\multicolumn{1}{|c|}{\small{ADS-B}} & \small{Aircraft broadcasts GPS-based position from transponder to ADS-B antenna} & \small{Aircraft position, altitude (more frequently and accurately), identity, ground speed, and flight information.} & \small{Susceptible to eavesdropping and falsification}\\ 
	\hline
  \end{tabular}
  \label{tab:atc_comm_techs}
\end{table*}
\fi

%DDS Network
\paragraph{Data Distribution Service (DDS) Network.} ATC position reporters generate various types of data to be provided to several end-systems. The Data Distribution Service (DDS) framework is a common option for this task and is used in a number of sectors including ATC~\cite{Chen2010,DDSFoundation}. DDS is a middleware connectivity standard where data sources (publishers) broadcast information to systems interested in this data (subscribers). The \name~architecture uses the DDS framework to aggregate all the PSR, SSR, and ADS-B data into a centralized network of standardized DDS packets.

\subsection{ADS-B Anomaly Detection}\label{sec_adsb_ad}
%Protocol modifications
One stream of research for preventing the transmission of falsified ADS-B packets has looked at securing the protocols through the use of encryption and authentication methods~\cite{Feng2010,FINKE20133,WU_2019,viggiano2001,Kim_8171273}. These approaches either require modifying the current specification of the ADS-B protocol or require additional equipment to be deployed in ATC infrastructure to certify the authenticity of ADS-B messages. Coordinating such schemes across the aviation industry is a daunting task and is not a near-term solution.

%Physical Detection (Doppler, RSS)
Non-cryptographic methods for detecting anomalous ADS-B packets aim to verify that the physical properties of received ADS-B messages are in-line with that of actual aircraft~\cite{schafer_7163027,Ghose_7311412,Schafer_2016_SMV,Strohmeier_2015_PHY}. These types of approaches verify that the Doppler shift measurements or the Received Signal Strength (RSS) of transmitted ADS-B packets match that of legitimate aircraft. These approaches do not require modifications to the ADS-B protocol, however, they assume an attacker does not have the ability to send falsified ADS-B messages which completely resemble those of legitimate aircraft. This assumption can be seen as flawed since it naively assumes the limits of an attacker's capabilities and does not prepare for the most sophisticated attacks.

%In practice this assumption is widely regarded as flawed, since it naively assumes the limits of an attacker's capabilities and does not prepare for the most sophisticated attacks.

%ML
Advances in Machine Learning (ML) have prompted researchers to investigate how to detect anomalous ADS-B messages which fall outside of established statistical bounds. Deep Neural Networks (DNNs) have been proposed to detect falsified ADS-B messages and falsified aircraft by training on labeled datasets of legitimate and malicious messages~\cite{ying19}. The use of Long Short-Term Memory (LSTM) has been proposed to learn trajectories of typical flights and trigger an alarm when a given flight path deviates from its normal path~\cite{Habler2018}. These approaches show the feasibility of detecting falsified ADS-B messages through ML pattern matching, however, they require the captured samples to cover enough statistical variety to prevent overfitting to particular patterns, which is difficult to collect in cybersecurity domains.

\subsection{Ontologies for Attack Detection}\label{sec_onto_attack_detection}

Ontologies have been used to model hierarchies of concepts in both academia and industry in a range of fields including cybersecurity~\cite{Souag2012,Arbanas2015} and aviation~\cite{Keller}. A key characteristic of ontologies is that all data is stored as nodes in a graph, where directed edges between nodes capture their relationship. Researchers have proposed to leverage this property for intrusion detection tasks~\cite{LI20107138,Sadighian2014,Coppolino2009}. These previous works demonstrate how attacks can be detected using ontologies in traditional Information Technology (IT) environments (e.g. computer networks). The premise being that as low-level alerts are generated from different sensors, such as an Intrusion Detection System (IDS) or a Security Information and Event Management (SIEM) tool, they can be translated into ontological instances and be stored in an ontological database. The ontological database is then queried at a regular interval to look for violations to normal operating criteria. The queries can be seen as a mechanism for aggregating the meaning of low-level alarms into more semantically rich detection rules. These rules put the occurrence of low-level alarms into some established context and is known as alert correlation.

%The machine-readable nature of ontologies allows for automated reasoning and querying capabilities upon security data (e.g. packet captures, system logs, etc.) represented as ontological instances. 

%To our knowledge, no method has been proposed to use ontologies in an attack detection framework  against ATC systems. This paper proposes an ontological approach for detecting malicious messages injected into ATC communications infrastructure. The details of this approach are provided in Section \ref{sec_onto_detection}.

%/textcolor{blue}{clarify the difference between GraphDB and Ontologies}

The most popular ontology language is the Web Ontology Language (OWL) \cite{OWL}, which utilizes data written in the Resource Description Framework (RDF) format~\cite{RDF}. RDF data is stored in an ontological database and is queried using the SPARQL language. Unlike relational databases, RDF databases are schema-less and allow a higher degree of flexibility when assigning properties to entities. In RDF, each statement is represented as a \textit{triple} consisting of a subject, a predicate, and an object. Thus, RDF triples can be conceptualized as a node and arc labeled graph. Each element of a triple is identifiable by a Uniform Resource Identifier (URI). The schema defining the ontological classes and the instantiations of these classes appear in the same repository as RDF triples. %There are different formats for representing RDF data such as RDF/XML, Turtle, JSON, and N-Triples. 

The advantage gained from detecting attacks through an ontological-based alert correlation framework is that there is greater explainability as to why an anomaly was flagged, since the relationship between low-level alarms is captured. Describing knowledge domains with ontologies goes beyond the capability of taxonomies as it defines the relationships between entities, allowing all information within the ontology to form a graph structure and be reasoned upon~\cite{Joshi2003}.

In a security setting, an ontology-based anomaly detection approach may have several benefits over machine learning. In machine learning, an anomaly detection agent learns a decision boundary based on the distribution of samples it has trained upon and is susceptible to generating false positives when it is provided unexpected samples to classify. However, by describing detection rules with ontologies, the attack space can potentially have more coverage with explainable rules.

%It was stated that one of the benefits of ontologies is the graph-based nature of capturing data. 
%In this paper we opted for ontologies based on the RDF standard, however, other graph data storage platforms exist, both open source and proprietary. The advantage of using RDF based ontologies is that they can easily be extended with other ontologies and that it is a W3C standardized data format.

%-- Section 3: ATC-Sense System Overview
\section{\name~System Overview}\label{sec:system_overview}
%This section outlines the threat model which \name~considers, provides the details of the \name~implementation, and describes the implemented detection constraints. The detection constraints use SPARQL queries to reason upon the track information, radar readings, and flight plans of aircraft traveling in a simulated airspace.
This section outlines the threat model which \name~considers, provides the details of the \name~implementation, and describes the implemented detection constraints. %The detection constraints use SPARQL queries to reason upon the track information, radar readings, and flight plans of aircraft traveling in a simulated airspace.

\subsection{Threat Model}\label{sec_threat_model}
%Prior to the 2000s the general public had very limited abilities for interfering with avionics communications. Adversaries were limited to militaries or nation-states with electronic warfare capabilities. Attacks were limited to analog technologies, which are easier to detect. Also, the level of insider knowledge related to communications systems and aviation conduct was not readily available to the public~\cite{Strohmeier2017}.

%However, as there is a shift towards digital technologies and automated processes, there is a much larger attack surface. The financial barrier to access SDR technologies is relatively low with competent transceivers starting around \$150 USD. Additionally, access to open-source software which can process avionic communications can be readily downloaded. The availability of aviation knowledge is also easily accessed over the Internet. Protocol specifications can be accessed, plane-tracking websites exist, flight plans are made public, and individuals can even capture real-world ADS-B communications with home-made antennas. 

\begin{figure}[h!]
\centering
	\includegraphics[scale=0.25]{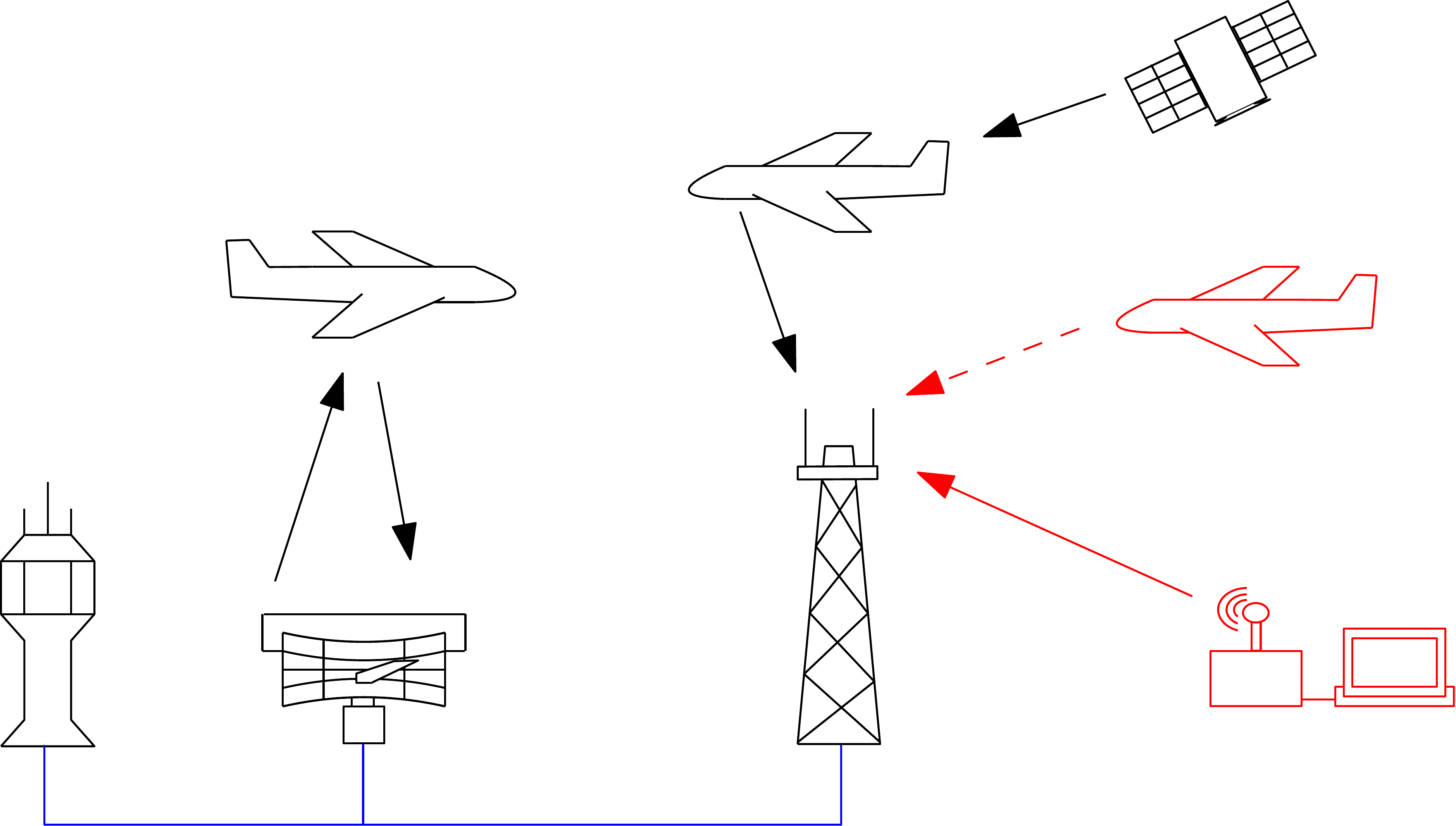}
	\caption{ATC Adversary Model. An attacker uses an SDR device to send falsified ADS-B messages to an ADS-B antenna. The reporters communicate with the ATC tower over a DDS network. Falsified aircraft appear to exist to ATC personnel.}
	\label{fig:adsbattack}
\end{figure}

\begin{figure}[h!]
%\begin{figure*}[h!]
\centering
	\includegraphics[width=\linewidth]{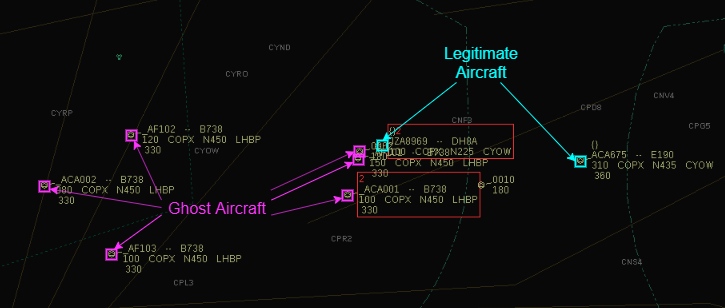}
	\caption{ATC display with multiple falsified (ghost) aircraft}
	\label{fig:ghostfloodingattack}
\end{figure}

This paper considers an attacker which employs an SDR device with spoofing capabilities. The attacker has the capability to send spoofed ADS-B messages to a targeted ADS-B antenna receiver within the range of the SDR device. We assume the attacker has a complete understanding of the ADS-B protocol, knowledge of reporter ground station locations, and access to real-world flight plans (freely available on the Internet). The SDR based attack is outlined in Figure~\ref{fig:adsbattack}. The effect of the attack on the ATC display is provided in Figure~\ref{fig:ghostfloodingattack}. \name~is conceived to find malicious ADS-B packets by connecting to a DDS network, converting the DDS-based ADS-B, SSR, and PSR data into RDF format, and using SPARQL queries to reason upon the RDF data to infer anomalies.

%and uses SPARQL queries to reason upon DDS-based ADS-B, SSR, and PSR traffic that is converted into RDF data.

%\name~works by analyzing all incoming ADS-B, SSR, and PSR communications to infer the presence of anomalous ADS-B messages. \name~is connected to a DDS network and converts all DDS-based ADS-B, SSR, and PSR traffic into RDF format to be queried in order to infer the presence of malicious packets.

%\end{figure*}

\subsection{System Implementation}\label{sec_implementation}
\begin{figure*}[h!]
\centering
	\includegraphics[width=\linewidth]{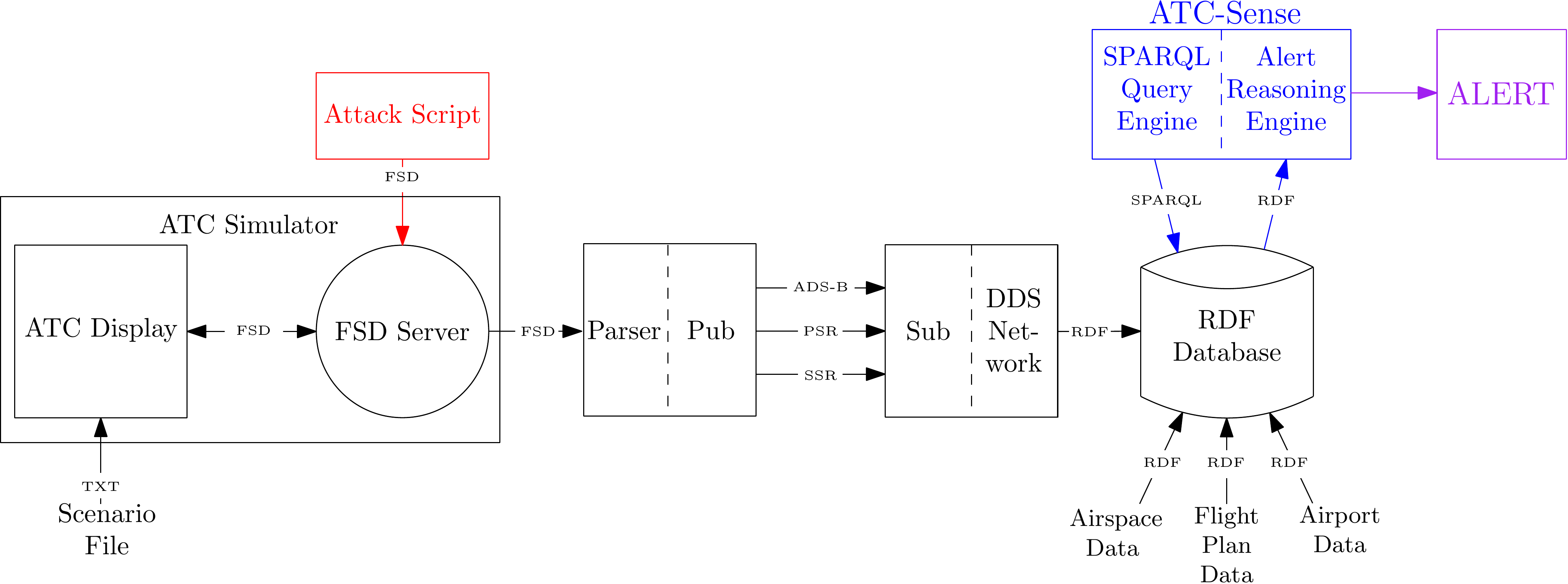}
	\caption{Block diagram of \name~in the experimental setup}
	\label{fig:implementation}
\end{figure*}

\name~was developed with the objective of detecting anomalous ADS-B messages in an operational ATC system. To achieve this, a simulated ATC environment is used to generate realistic ATC related network traffic and is integrated with \name. In this setup, an attack is generated by injecting false ADS-B communications traffic into the ATC simulation with the intent of misleading the ATC controller. All communication packets (PSR, SSR, and ADS-B) are converted into RDF format and are reasoned upon by an ontology-based query system to detect malicious input. An overview of the implementation is provided in Figure~\ref{fig:implementation}.

The \textbf{ATC Simulator} is a freely available Windows application called Euroscope which simulates the transmission of actual communications packets between aircraft and ground towers, issues automated controller commands to simulated aircraft, and provides a graphical display to observe the scenario from the viewpoint of a human controller. Euroscope is the most popular ATC client software used by the Virtual Air Traffic Simulation Network (VATSIM), an online community of over 79,000 active members of hobbyist virtual pilots and air traffic controllers which conduct real-time virtual flights between real-world airports~\cite{VAT}.

Euroscope is comprised of two central components, the \textbf{ATC Display} and the \textbf{FSD Server}. Euroscope exchanges FSD Packets between the FSD Server, performing the commands of the ATC simulation, and the ATC Display which displays the events occurring in the simulation (e.g. aircraft positions, flight paths, squawk codes, etc.). A \textbf{Scenario File} is loaded into the ATC Display, prior to executing a simulation, to import airport features, physical barriers, and aircraft positions into the simulation. In Euroscope's simulation mode, the FSD server will automatically issue ATC commands to the aircraft and guide them to the airport based on an internal control algorithm. The \textbf{Attack Script} is a custom Python script which sends malicious packets to the FSD Server which are meant to mislead the control algorithm.

In real-world ATC systems, FSD packets are not used between ground towers and the aircraft, instead data is transferred using radar signals (PSR, SSR) and ADS-B packets. We have configured the FSD Server to send all the FSD packets generated within Euroscope to a custom \textbf{Parser} program written in C, to convert the FSD packets into their corresponding PSR, SSR, and ADS-B signals. Following the Pub-Sub pattern, all the generated  PSR, SSR, and ADS-B signals are published by the \textbf{Pub} program to the \textbf{Sub} component, which then are made available on the \textbf{DDS Network}, all of which is done using Python scripts. This method allows for the generation of DDS Network traffic in-line with what is used by actual ATC systems. 

As the DDS Network receives PSR, SSR, and ADS-B signals, a Python script is used to convert the data into RDF triples and passes the RDF data to the \textbf{RDF Database}. The database tool used is a W3C compliant RDF triplesore called GraphDB~\cite{GDB}, which has a base free-to-use license. GraphDB operates as a web server which can insert RDF triples and perform SPARQL queries through a REST API. Prior to executing a simulation, the RDF Database is populated with some static base information which includes \textbf{Airspace}, \textbf{Flight Plan}, and \textbf{Airport Data}. This base information is used in conjunction with the real-time data coming from the DDS Network in the reasoning process to infer the presence of malicious input.

\name~is used to determine the presence of anomalous ADS-B packets and is comprised of the \textbf{SPARQL Query Engine} and the \textbf{Alert Reasoning Engine}. The SPARQL Query Engine is a series of Python scripts which queries the RDF Database at a regular interval, over a REST API, to retrieve ATC related entities that exist in the database. The Alert Reasoning Engine is a Python script which performs if-else type logic on the retrieved RDF data to infer the presence of anomalous packets. An \textbf{Alert} is provided if an anomaly is detected. The alert indicates which packet is anomalous, along with its reported coordinates. This information could be used to provide a visual alert to a controller on a ATC display screen. This entire architecture has been implemented to execute on a single 64-bit Kali Linux machine running on an Intel Core i7-3770 processor with 4 cores of 3.40GHz. 

\subsection{Track Constraints-Based Detection: Consistent Origin}\label{track_detection}
The track is the series of reported positions associated to an aircraft and is comprised of timestamped PSR, SSR, and ADS-B reports. Should a ghost aircraft be injected, it may not have a logical starting location. A single ghost aircraft may go unnoticed to a controller, additionally, multiple ghost aircraft could potentially flood the controller's screen. This detection logic verifies if all newly created tracks have a logical starting point. The SPARQL query in Listing~\ref{lst:track_query} is used to select the first ADS-B position report of each newly created track. 

%This may go unnoticed to an overwhelmed controller and cause the controller to issue incorrect commands to pilots. %Also, multiple ghost aircraft could completely flood the screen and severely harm the controller's abilities.

\begin{lstlisting}[caption=Select First ADS-B Position Report of the Track, label={lst:track_query}, captionpos=b]
<@\begin{scriptsize}\textcolor{codegreen}{PREFIX} \textcolor{eclipseBlue}{atc-adsb}: \textcolor{codeyellow}{<http://atcs.ex/atc/dds-topics/adsb-broadcast\#>}\end{scriptsize}@>
<@\begin{scriptsize}\textcolor{codegreen}{PREFIX} \textcolor{eclipseBlue}{atc-adsb}: \textcolor{codeyellow}{<http://atcs.ex/atc/atc-core\#>}\end{scriptsize}@>
<@\begin{scriptsize}\textcolor{codegreen}{PREFIX} \textcolor{eclipseBlue}{atc-data}: \textcolor{codeyellow}{<http://atcs.ex/atc/atc-data\#>}\end{scriptsize}@>
SELECT ?report ?lat ?long ?alt ?eID ?call WHERE {
{?report	<@\textcolor{codegreen}{a}@>	<@\textcolor{eclipseBlue}{atc-adsb}:\textcolor{codegreen}{ADSBFlightPosition}@>;
					<@\textcolor{eclipseBlue}{atc-core}:\textcolor{codegreen}{hasTrackRank}@>   ?rank;
					<@\textcolor{eclipseBlue}{atc-adsb}:\textcolor{codegreen}{hasLatitude}@>    ?lat;
					<@\textcolor{eclipseBlue}{atc-adsb}:\textcolor{codegreen}{hasCallsign}@>    ?call;
					<@\textcolor{eclipseBlue}{atc-adsb}:\textcolor{codegreen}{hasLongitude}\hspace{0.9pt}@>   ?long;
					<@\textcolor{eclipseBlue}{atc-adsb}:\textcolor{codegreen}{hasAltitude}@>    ?alt;
					<@\textcolor{eclipseBlue}{atc-adsb}:\textcolor{codegreen}{hasEquipmentID}\hspace{2.0pt}@> ?eID.}
FILTER(?rank=1)
}
\end{lstlisting}

This SPARQL query returns all of the ADSB Flight Position Reports where the rank is equal to 1. The rank is used to determine the order of processed position reports for a given aircraft. The returned ADS-B Flight Position Reports have their associated latitude and longitude coordinates, thus constitute a point. The distance of this ADS-B point between nearby airports and the border of the ADS-B coverage area is calculated. If this report did not appear next to the border of an ADS-B coverage range or near an airport, then the associated ADS-B packet can be flagged as anomalous along with its coordinate position. %SPARQL is not optimized to perform certain mathematical operations, such as calculating distance between different points, therefore we use Python scripts to execute the SPARQL queries to retrieve the ADS-B position reports, then calculate the distances using Python functions. 

\subsection{Radar Constraints-Based Detection: Reporter Consistency}\label{radar_detection}
An attacker can falsify ADS-B messages, however, they cannot falsify PSR readings which operate according to physical properties. Within this constraint logic, SPARQL queries determine if the reported ADS-B positions are associated with valid PSR radar positions. A linear interpolation between the trajectories of the received PSR and SSR tracks is done with the ADS-B tracks to match similar tracks. If an ADS-B track is not associated to a PSR or SSR track, then it is flagged as anomalous. The SPARQL query in Listing~\ref{lst:radar_query} selects all the ADS-B Flight Position Reports which are not associated to any PSR or SSR track. The selected ADS-B Flight Position Reports are flagged as anomalous if they appear within range of a PSR or SSR radar.

%\name~can detect anomalous ADS-B packets by verifying that all received ADS-B packets within the PSR or SSR coverage range have an associated PSR or SSR track. 

\begin{lstlisting}[caption=Select ADS-B Position Reports with no associated PSR Track, label={lst:radar_query}, captionpos=b]
<@\begin{scriptsize}\textcolor{codegreen}{PREFIX} \textcolor{eclipseBlue}{atc-adsb}: \textcolor{codeyellow}{<http://atcs.ex/atc/dds-topics/adsb-broadcast\#>}\end{scriptsize}@>
<@\begin{scriptsize}\textcolor{codegreen}{PREFIX} \textcolor{eclipseBlue}{atc-adsb}: \textcolor{codeyellow}{<http://atcs.ex/atc/atc-core\#>}\end{scriptsize}@>
SELECT ?track ?report ?lat ?long ?time WHERE {
{?report	<@\textcolor{codegreen}{a}@>	<@\textcolor{eclipseBlue}{atc-adsb}:\textcolor{codegreen}{ADSBFlightPosition};@>
					<@\textcolor{eclipseBlue}{atc-core}:\textcolor{codegreen}{hasTrackRank}@>          ?rank;
					<@\textcolor{eclipseBlue}{atc-core}:\textcolor{codegreen}{isAssociatedWithTrack}\hspace{1.5pt}@>  ?track;
					<@\textcolor{eclipseBlue}{atc-adsb}:\textcolor{codegreen}{hasLatitude}@>           ?lat;
					<@\textcolor{eclipseBlue}{atc-adsb}:\textcolor{codegreen}{hasLongitude}@>          ?long;
					<@\textcolor{eclipseBlue}{atc-adsb}:\textcolor{codegreen}{hasTimeStamp}@>          ?time.}
MINUS {?track <@\textcolor{eclipseBlue}{atc-core}:\textcolor{codegreen}{hasSimilarTrack}@> ?tk}
}ORDER BY ?track ASC(?time)
\end{lstlisting}

\subsection{Flight Constraints-Based Detection: Existence of Flight Plan}\label{flight_constraints}
With this detection constraint, all ADS-B tracks are verified against actual flight plans that have been loaded into the ontological database. If an ADS-B track does not have the properties of a registered flight plan then it's associated packets are flagged as anomalous. The SPARQL query in Listing~\ref{lst:flight_query} selects all ADS-B reports which do not have a call sign associated to a flight plan. %These selected ADS-B Flight Position Reports are flagged as anomalous since they do not have a flight plan. 

\begin{lstlisting}[caption=Select ADS-B Position Reports with invalid callsign, label={lst:flight_query}, captionpos=b]
<@\begin{scriptsize}\textcolor{codegreen}{PREFIX} \textcolor{eclipseBlue}{atc-adsb}: \textcolor{codeyellow}{<http://atcs.ex/atc/dds-topics/adsb-broadcast\#>}\end{scriptsize}@>
<@\begin{scriptsize}\textcolor{codegreen}{PREFIX} \textcolor{eclipseBlue}{atc-adsb}: \textcolor{codeyellow}{<http://atcs.ex/atc/atc-core\#>}\end{scriptsize}@>
SELECT ?callsign ?report ?lat ?long ?time WHERE {
{?report	<@\textcolor{codegreen}{a}@>	<@\textcolor{eclipseBlue}{atc-adsb}:\textcolor{codegreen}{ADSBFlightPosition}@>;
					<@\textcolor{eclipseBlue}{atc-adsb}:\textcolor{codegreen}{hasCallsign}@>  ?callsign;
					<@\textcolor{eclipseBlue}{atc-adsb}:\textcolor{codegreen}{hasLatitude}@>  ?lat;
					<@\textcolor{eclipseBlue}{atc-adsb}:\textcolor{codegreen}{hasLongitude}@> ?long;
					<@\textcolor{eclipseBlue}{atc-adsb}:\textcolor{codegreen}{hasTimeStamp}@> ?time.}
MINUS {?fp <@\textcolor{eclipseBlue}{atc-core}\textcolor{codegreen}{:hasCallsign}@> ?callsign}
\end{lstlisting}

%-- Section 4: Evaluation
\section{Evaluation}\label{sec:evaluation}
This section explains the simulation scenario used to perform our experiments, demonstrates the computational performance metrics gathered from the experiments, and provides an analysis of the \name~detection approach.
 
\subsection{Simulation Scenario}
We evaluate \name~by using a fixed simulation scenario with 10 legitimate aircraft and attack the scenario by varying the amount of ghost aircraft from 0 to 5. \name~is able to detect all of the ghost aircraft injected into this scenario and demonstrates that ontology-based detection is an applicable approach in this domain. Since we detect all of the ghost aircraft, we don't analyze detection statistics. Instead, we examine the computational metrics of \name~to gain insight into the resources required to perform ontology-based anomaly detection and to identify where future work is needed to scale this approach to real-world settings.

The simulation scenario runs for a total of 400 seconds (nearly 7 minutes). The first 3 minutes involve the initialization of the simulation. Within this initial period, the 10 legitimate aircraft operate within the simulation and there are no ghost aircraft. No packets are sent to GraphDB during this period and are instead stored in a buffer. Once 400 DDS packets are generated by the simulation, around the 3 minute mark, the packets are converted to RDF and inserted into GraphDB. At this point the simulation is initialized with the normal aviation entities. After this initialization, batches of 50 packets are inserted into GraphDB as they are generated by the simulation. We use these batches because single DDS packets cannot be inserted into GraphDB as RDF in real-time as they arrive, since the insertion does not complete in time before another insertion request arrives and is consequently dropped by GraphDB.

% (discussed in Section~\ref{sec:discussion}). 

The ghost aircraft are injected into the simulation at the 3 minute mark. Simulations are run with attacks ranging from 0 to 5 ghost aircraft which can be detected by our constraints logic. During the entire 400 seconds of the simulation one of the three detection constraints logic is operating and is actively looking for anomalous ADS-B packets. For each of the 3 detection constraints, 10 simulations are run with attacks ranging from 0 to 5 ghost aircraft. All results are averaged over the 10 runs. In total $3*6*10=180$ simulations were recorded, resulting in $(180~ \text{sims}*400~\text{secs}) / 3600~\text{secs} = 20$  hours of total simulation time.

\begin{figure}[h!]
\centering
	\includegraphics[scale=0.20]{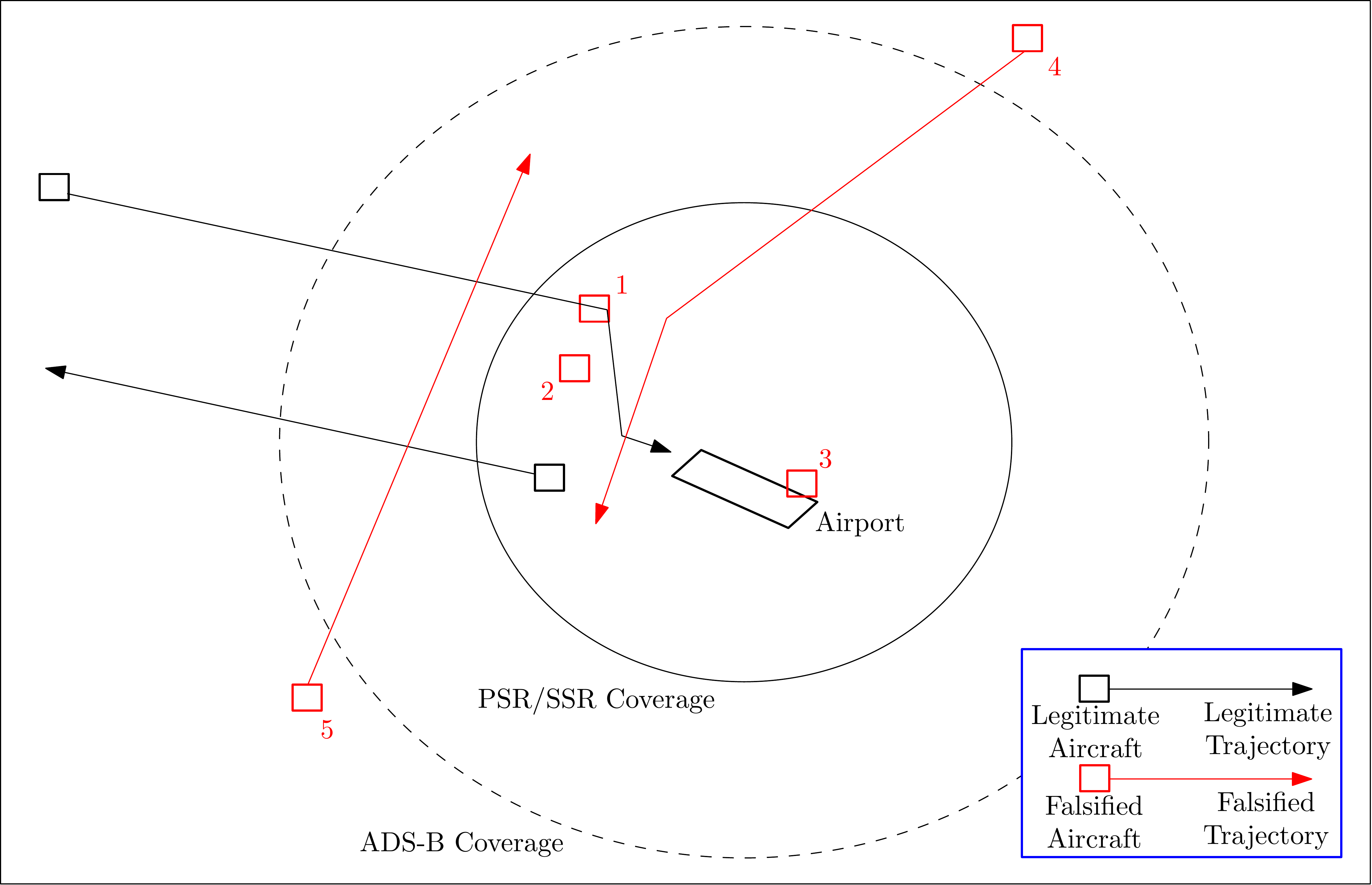}
	\caption{Simplified representation of the simulation scenario with 5 falsified aircraft being injected. (1) \& (2) Static ghost aircraft, (3) Static ghost aircraft near an airport, (4) Moving ghost aircraft which travels from ADS-B coverage area into PSR/SSR/ADS-B coverage area, (5) Moving ghost aircraft which travels solely in ADS-B coverage area.}
	\label{fig:scenario_experiments_1}
\end{figure}

Figure~\ref{fig:scenario_experiments_1} shows a simplified representation of the simulation scenario. This figure is meant to convey the behavior of the injected ghost aircraft and does not reflect all of the 10 legitimate aircraft. The ghost aircraft are labeled from (1) to (5). The falsified aircraft labeled (1) and (2) are static ghost aircraft, (3) is a static ghost aircraft near an airport, (4) is a moving ghost aircraft which travels from the ADS-B coverage area into the PSR/SSR/ADS-B coverage area, and (5) is a moving ghost aircraft which travels solely in the ADS-B coverage area.

The Track Constraint for Consistent Origin detects the falsified aircraft labeled (1) and (2), since these aircraft appear at invalid locations. This is the fastest constraint to check and would be used first in a deployed system. The Radar Constraint for Reporter Consistency detects the falsified aircraft labeled (1)-(4). The falsified aircraft labeled (1)-(3) are detected because they have a stationary track. The falsified aircraft labeled (4) is detected because there is no PSR/SSR track associated to the ADS-B track. The Flight Constraint for Existence of Flight Plan detects all the falsified aircraft. In this simulation none of the falsified aircraft have an associated Flight Plan. 

\subsection{Performance Metrics}

We measure the performance of the detection approach by analyzing how the computational overhead of the detection process is affected with an increased workload. This is to address the research goal of understanding the feasibility of using an ontology-based anomaly detection approach in a real-time setting.

\begin{figure*}[h!]
\centering
	\includegraphics[width=\linewidth]{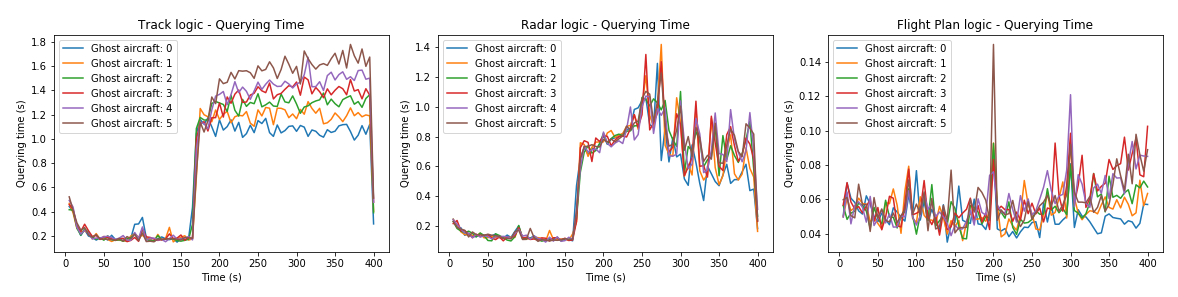}
	\caption{Execution time of SPARQL queries (average of 10 simulations)}
	\label{fig:10_time}
\end{figure*}

\paragraph{SPARQL Query Time.} This is the amount of time needed to execute the SPARQL queries. Note that this quantity is the sum of the insert and select operating times. Figure~\ref{fig:10_time} shows the querying time during the simulations for the implemented Track, Radar, and Flight Plan constraints logic, recorded every 5 seconds. 
The query logic is operating for the entire 400 seconds of the simulation and the first insert occurs at 180 seconds. The initial 180 seconds have very low time since the detecion queries are being performed with no aviation entities in GraphDB. At 180 seconds there is a large increase in query time when the initial 400 packets are inserted. After 180 seconds there are noticeable spikes at some recurring interval. This corresponds to the batches of 50 packets being inserted into GraphDB. The query time increases at roughly a constant rate relative to the number of ghost aircraft for each of the Track, Radar, and Flight Plan constraints. The Track query logic is the most complex with the highest query time, the Radar query logic is slightly less complex with a reduced query time, and the Flight Plan logic is considerably less complex with a much lower query time.

\begin{figure*}[h!]
\centering
	\includegraphics[width=\linewidth]{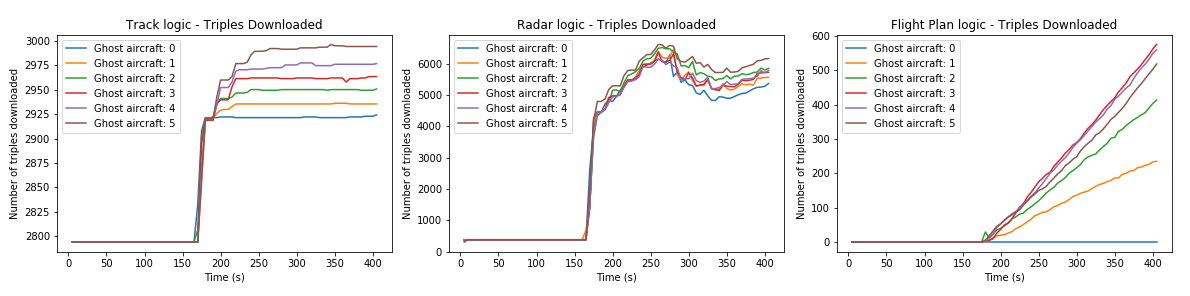}
	\caption{Number of RDF triples downloaded/queried (average of 10 simulations)}
	\label{fig:20_triples}
\end{figure*}

\begin{figure*}[h!]
\centering
	\includegraphics[width=\linewidth]{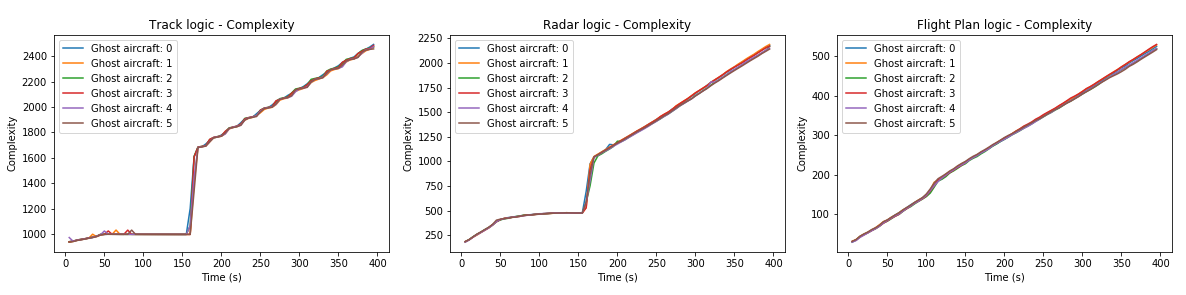}
	\caption{Complexity of SPARQL queries (average of 10 simulations)}
	\label{fig:30_complexity}
\end{figure*}

\paragraph{RDF Triples Downloaded.} This is the number of RDF triples downloaded from GraphDB by performing SPARQL select operations. Figure~\ref{fig:20_triples} shows the number of triples downloaded during the simulations for the implemented Track, Radar, and Flight Plan constraints logic, recorded every 5 seconds. In the initial 180 seconds there are no packets queried from GraphDB during the initialization period. For the Track and Radar logic the number of triples downloaded increases roughly to a constant rate relative to the number of ghost aircraft. In the Flight Plan logic we do see a large linear increase, however, the total number of triples is relatively low and is an order of magnitude smaller than the other constraints.

\begin{table*}[t]
\centering
\caption{GraphDB Read/s and Write/s for the detection constraints (average of 10 simulations)}
\begin{tabular}{|c||c|c||c|c||c|c|}
\hline
 & \multicolumn{2}{|c||}{\textbf{Track Constraints}} & \multicolumn{2}{|c||}{\textbf{Radar Constraints}} & \multicolumn{2}{|c|}{\textbf{Flight Constraints}} \\
\hline
\textbf{\# Ghosts} & \textbf{Reads/s} & \textbf{Writes/s} & \textbf{Reads/s} & \textbf{Writes/s} & \textbf{Reads/s} & \textbf{Writes/s} \\
\hline
0 & 502.5751 & 5.3093 & 303.3250 & 5.3988 & 15.2259 & 4.9518\\
1 & 512.6787 & 5.3361 & 307.4927 & 5.3689 & 22.2010 & 5.5481\\
2 & 512.2853 & 5.2852 & 308.0748 & 5.2636 & 25.4083 & 5.3126\\
3 & 514.2179 & 5.3939 & 309.2526 & 5.2867 & 29.5177 & 5.3133\\
4 & 518.3481 & 5.2978 & 309.5447 & 5.3408 & 30.0592 & 5.5255\\
5 & 513.5878 & 5.4505 & 316.4494 & 5.4167 & 29.4796 & 5.4622\\
\hline
\end{tabular}	
	\label{table:readwrite}
\end{table*}

\paragraph{SPARQL Complexity.} This is a quantity which represents the estimated number of iterations required to perform the SPARQL queries. Figure~\ref{fig:30_complexity} shows the complexity of the SPARQL queries during the simulations for the implemented Track, Radar, and Flight Plan constraints logic, recorded every 5 seconds. In the initial 180 seconds the queries are less complex since the initialization data has not been inserted into GraphDB. There is a noticeable jump in complexity after the initial packets are inserted into GraphDB. In all three cases the complexity of the queries is marginally affected by increasing the number of ghost aircrft. However, there is a dramatic linearly increasing total complexity as the simulation continues. This occurs because as more triples are inserted into the database the queries need to iterate over an increasingly large number of entities. 

\paragraph{GraphDB Read/s and Write/s.} GraphDB provides the average number of RDF reads and writes per second. Table~\ref{table:readwrite} shows the reads/s and write/s when using each of the detection constraints. For each of the constraints the reads/s tends to gradually increase with the number of ghost aircraft. The time of the Reads/s is proportional to the complexity of the SPARQL queries in each of the detection constraints. The writes/s is not affected by the number of ghost planes nor by which detection constraints are used. This is likely due to a limitation of GraphDB running in our setup.

\subsection{Analysis and Future Work}
The experiments presented in this work are an initial study of the feasibility of using an ontology-based detection approach for detecting anomalous ADS-B messages in a real-time security setting. Previous security related research use ontology-based frameworks for detecting anomalies in static datasets, while this work is the first, to the authors' knowledge, to do so in a real-time setting. The purpose of measuring the computational overhead of the  approach is to gain an understanding of issues that may arise when scaling real-time ontology-based anomaly detection to real-world applications.

The SPARQL query times generally increase at a constant rate as more entities are present in the simulation. This is particularly noticeable with the Track Logic which shows that when there are no ghost aircraft the execution time is roughly 1.2 seconds and while there are 5 ghost aircraft the execution time jumps to around 1.6 seconds. In such a small setting this is not drastic, however, if there are to be potentially tens or hundreds of entities, the SPARQL execution times will likely be magnitudes higher and may become too slow for a time-critical detection approach. Additionally, the complexity of the SPARQL queries demonstrates a dramatic linear increase as the simulation executes. The RDF triples are entering the database at a relatively constant rate, forcing the SPARQL queries to iterate over a steadily-increasing set of data. This observation hints at a need to further analyze at what point this increased complexity becomes a bottleneck for detecting anomalies. A primary conclusion from this analysis is that work needs to be done to speed up the execution times of the queries and limit their computational complexity. A stream of future work could be to devise a process for removing existing RDF triples from an ontological database that are deemed to no longer be necessary in the detection process, in hopes of reducing the computational overhead of the SPARQL queries.

Interpreting the reads/s and writes/s also offers some valuable insights. There is likely some upper limit of reads/s and writes/s, for a particular architectural setup. When given an environment with a large number of entities generating position reports, it is conceivable that not enough RDF data can be inserted and reasoned upon quick enough to perform adequate threat detection reasoning. Future work should strive to formalize an understanding of how many entities can be reasoned upon in an adequate time, for a particular architectural setup.

It must also be mentioned that the defensive coverage of the implemented detection constraints, at the time of writing, is relatively low. Particular scenarios can be constructed which would not be detected. Consider the scenario where a ghost aircraft enters the ADS-B coverage range at a time very close to that of an actual registered aircraft. Our constraints will not be able to distinguish between the legitimate aircraft and the ghost aircraft because the track reports will be coming from similar locations. An additional set of physics-based constraints should be used to determine that the trajectories of reported ADS-B positions follow the laws of physics.

\section{Conclusion}\label{sec:conclusion}
This paper presents \name, an ontology-based ADS-B anomaly detection tool. This tool has been integrated with a simulated ATC environment which can be attacked with falsified ADS-B messages. We developed an ATC ontology to reflect the entities and operating procedures of aviation environments in a formalized language. Semantic queries based on track, radar, and flight plan constraints are used to detect violations to aviation logic caused by malicious ADS-B messages injected by a capable adversary. Simulated experiments with falsified ghost aircraft operating under various scenarios demonstrate the feasibility of ontology-based anomaly detection for ATC systems and other real-time security domains. Future work involves increasing the defensive coverage of the proposed ATC detection rules and finding methods to scale this approach to larger settings.

%This approach opens the door for community validation by performing attack detection using larger scale simulations, such as in real online ATC communities. 

\bibliographystyle{vancouver}
\bibliography{atc_paper_semantics}

%\begin{thebibliography}{99}

%\bibitem{r1}
%Petitti DB, Crooks VC, Buckwalter JG, Chiu V. Blood pressure levels before dementia.
%Arch Neurol. 2005 Jan;62(1):112-6.

%\bibitem{r2}
%Rice AS, Farquhar-Smith WP, Bridges D, Brooks JW. Canabinoids and pain. In: Dostorovsky JO,
%Carr DB, Koltzenburg M, editors. Proceedings of the 10th World Congress on Pain;  2002 Aug
%17-22; San Diego, CA. Seattle (WA): IASP Press; c2003. p. 437-68.

%\end{thebibliography}
\end{document}